\documentclass[
preprint,
 amsmath,amssymb,
 aps,
]{revtex4-1}

\usepackage{graphicx}
\usepackage{dcolumn}
\usepackage{bm}

\begin{document}


\title{Prediction of citation dynamics of individual papers\\}

\author{Michael Golosovsky}
\email{michael.golosovsky@mail.huji.ac.il}
\affiliation{The Racah Institute of Physics, The Hebrew University of Jerusalem, 9190401 Jerusalem, Israel\\
}%
\date{\today}
\begin{abstract}
We apply  stochastic model of citation dynamics of individual papers developed in our previous work (M. Golosovsky and S. Solomon, Phys. Rev. E\textbf{ 95}, 012324 (2017)) to forecast  citation career  of individual papers.  We focus not only on the estimate of the future citations of a paper  but on the probabilistic margins of such estimate as well.
\begin{description}
\item[PACS numbers]01.75.+m, 02.50.Ey, 89.75.Fb, 89.75.Hc
\end{description}
\end{abstract}
\maketitle
\section{Introduction}
The interest in predicting citation behavior of scientific papers is motivated by the need to forecast the journal impact factor, for  early identification of the  breakthrough  papers,  and career considerations \cite{Clauset2017,Zeng2017,Tahamtan2016}). Prediction is usually based on \emph{a priori}  and \emph{a posteriori} factors that, in principle, can determine citation career of a paper. The former factors are set at the moment of publication and these are  subject, title, author's previous record and reputation \cite{Castillo2007,Ke2013,Peters1994,Stegehuis2015,YanHuang2012,YanTang2011}, venue (journal) \cite{Lariviere2010,Didegah2013}, the length and the composition of the reference list \cite{Uzzi2013,YanHuang2012,Didegah2013,Klimek2016}, the style of the paper \cite{Letchford2015,Didegah2013,Fox2019}, etc.   The difficulty  of this approach is  that the most important attributes, such as novelty, originality, significance, and timeliness of the results  are qualitative. In principle, some of them can be quantified but this is challenging. A brilliant example of such quantification is  Ref. \cite{Uzzi2013} which managed to characterize the novelty of a paper through diversity (frequency of atypical combinations) of its references. \emph{A posteriori}  factors develop during short time after the paper has been published  and these include the "impact factor"- the number of citations during a short period after publication \cite{Wang2013,Castillo2007,Adams2005,LiTong2015,Cao2016}, and the place that the paper occupies in its community.  There are two complementary approaches to  predict citation career of a paper basing on these factors.

Computer scientists focus more on  \emph{a priori } factors. They take  a large set of papers whose citation career has been evolving for a long time and use it for training, namely, they measure correlation between these factors and the number of citations of a paper in the long time limit.   Then, the factors are ranked according to their importance and  predictive model is built by machine learning. The general consensus is that predictive algorithm  shall use several factors or combination of them \cite{Bollen2009,Stegehuis2015}, whereas the relative weight of these factors for different disciplines can vary. It has been also realized that linear correlations do not tell the whole story \cite{Golosovsky2017,Wang2013,Castillo2007,LiTong2015}  and  predictive algorithm  shall be better nonlinear, similar to  that of Ref. \cite{LiTong2015}.  When the predictive algorithm has been validated, it works as follows. For a new paper, one determines   all relevant factors and builds a prediction. The result of prediction is the number of citations of a paper after some predetermined time. Although this prediction is probabilistic, the margins of predictability  were never studied properly.

The approach of  researchers with the background in natural sciences is different. They focus more on \emph{a posteriori} factors, such as recent citation history of a paper. They construct  empirical  models of citation dynamics which are based on some predetermined scheme of the citation process, namely, they assume a certain strategy that the author of a new paper adopts when he cites the previous studies. This model predicts a future citation behavior of a paper basing on its citation history and several  paper-specific parameters, the most important of them being fitness, a hidden parameter  that can be reliably estimated only after citation career of the paper has been developing for 2-3 years \cite{Hazouglou2017}. When the model has been constructed and validated, the prediction is performed as follows. One takes a new paper and, by studying its initial citation history, makes a probabilistic estimate of its fitness and other specific parameters. After such estimate has been made and the corresponding  parameters have been substituted into the model of citation dynamics, it predicts the number of citations of this paper in the long time limit. This approach has been most completely embodied in the Wang-Song-Barabasi model \cite{Wang2013}.

Bibliometric  analysis  considers both \emph{a priori} and \emph{a posteriori} factors. The researchers in this area  have long recognized that the early citation history of a paper is a good predictor of its future success.   On another hand, they were the first to draw attention to  sleeping beauties \cite{Glanzel2003,Raan2004,Ke2015,Wang2013}, the papers that  started to gain popularity long after publication.  Many important papers exhibited  the sleeping beauty behavior which no model of citation dynamics can predict.  Thus, the presence of such papers sets a limit to prediction of the future citation count of a paper. On another hand, this poor predictability is what makes  science   fun for so many researchers.

Our purpose is  to forecast the future citation career of a paper basing on our recently developed stochastic model of citation dynamics \cite{Golosovsky2017} .  This model includes several empirical parameters, some of them  are common to the whole discipline while all individual attributes of the paper are lumped into one parameter- fitness which does not vary with time. Our first goal is to explore the limits of predictability of the citation career of a paper with a given fitness, the uncertainty of prediction being related to intrinsic stochasticity of the citation process. Our second goal is to quantify the ingredients of fitness, in particular, we show how one can quantify such  attribute of a paper as timeliness of results.

\section{Stochastic model of citation dynamics- a summary}
 Assume a paper $j$ published in year $t_{j}$.  To quantify its citation dynamics, we introduce $\Delta K_{j}=k_{j}(t_{j},t_{i})dt$, the number of citations garnered by  this paper in the time window $(t_{i},t_{i}+dt)$ where $k_{j}(t_{j},t_{i})$ is the  paper's $j$ citation rate in year $t_{i}$. The model assumes that $\Delta K_{j}$ is a random variable that follows a time-inhomogeneous stochastic point process, namely, the probability of having  $\Delta K_{j}$ citations in a short time interval $dt$ is $\frac{\lambda_{j}^{\Delta K_{j}}}{\Delta K_{j}!}e^{-\lambda_{j}}$ where $\lambda_{j}dt$ is the paper-specific probabilistic citation rate. The model assumes that this rate  consists of the direct and indirect contributions,
\begin{equation}
\lambda_{j}(t_{j},t_{i})=\lambda_{j}^{dir}(t_{j},t_{i})+\lambda_{j}^{indir}(t_{j},t_{i}),
\label{paper-all-0}
\end{equation}
where the first term captures those  papers that cite paper \emph{j} and does not cite any other paper that cites \emph{j}; while the second term captures the papers that cite both \emph{j} and one or more of its citing papers.

The model yields the following expression for $\lambda_{j}(t_{j},t_{i})$
\begin{equation}
\lambda_{j}(t)=\eta_{j}R_{0}\tilde{A}(t)+
\int_{0}^{t}m(t-\tau)\frac{T(t-\tau)}{R_{0}}k_{j}(\tau)d\tau,
\label{4-citation5}
\end{equation}
where, in order to shorten notation, we introduced $t=t_{i}-t_{j}$, the number of years after publication, and dropped $t_{j}$.  The first addend in Eq. \ref{4-citation5} stays for the direct citation rate. Here, $\eta_{j}$ is the fitness of the paper $j$, $R_{0}$ is the average length of the reference list length of the papers published in year $t_{j}$, and $\tilde{A}(t)$ is the aging function.  The second addend  in  Eq. \ref{4-citation5} captures the indirect citation rate. Here, $m(t)$ is the average citation rate of the papers  published in  year $t_{j}$, $T(t)$ is the obsolescence function, and  $k_{j}(\tau)$ is the past citation rate of the paper $j$.


Our measurements yielded  the factors and functions, $R_{0},\tilde{A}(t),m(t)$ and $T(t)$. We shown that they  are the same for all papers in the same field published in the same year. Thus, the paper's individuality is captured by the fitness $\eta_{j}$  and by its past citation history, $k_{j}(\tau)$. To predict citation dynamics of the paper, we need to  measure its past citation dynamics and to estimate its fitness (which is supposed to be constant during paper's lifetime). Then we substitute these numbers into Eq. \ref{4-citation5} and run numerical simulation with known functions $R_{0},\tilde{A}(t),m(t)$ and $T(t)$.  Technically, Eq. \ref{4-citation5} describes a self-exciting or Hawkes process, since there is a positive feedback between the past and present citation citation rate. Hence, the prediction of future citations  is inherently probabilistic and its margins increase with time.

\section{Probabilistic character of the citation process. Implications with respect to predictability of future citations}
Citation process is stochastic, the stochasticity imposes limits on the predictability of future citations. Moreover, as we showed earlier \cite{Golosovsky2017}, citation dynamic of a paper follows a self-exciting (Hawkes) process whereby past fluctuations are amplified.  The positive feedback between past fluctuations and future citations renders the task of long-term prediction of citation behavior of a paper almost futile and limits predictive algorithms to the  range of 2-3 years.  In our previous study \cite{Golosovsky2012a} we illustrated this by measurements.

We explore here the following question: if we had known paper's fitness - what are the margins of predictability of its citation trajectory? To answer this question, we analyzed the relation between the paper's fitness and the number of citations it garners in the long-time limit. This was done using our calibrated and verified model of citation dynamics.  We wish to estimate, $K^{\infty}(\eta)$, the expected number of citations after 25 years for the paper with a certain fitness $\eta$. To this end, we performed numerical simulations based on Eq. \ref{4-citation5} with parameters for Physics papers published in 1984.  We considered 4000 papers  with the same fitness $\eta$, found statistical distribution of their citations  after 25 years, and measured the mean $K^{\infty}$  and the width of this  distribution.  We consider $K^{\infty}$ as the expected number of citations in the long time limit. 

Figure \ref{fig:fitness-bound}a shows that the expected number of citations, $K^{\infty}$, grows  nonlinearly  with  fitness $\eta$. Figure \ref{fig:fitness-bound}b focuses on the width of the $K^{\infty}$- distribution. We observe that for the papers with low $\eta$,   citation distribution in the long time limit  is wide, while for the papers with high $\eta$,  citation distribution in the long time limit  is narrow.  This means that while citation dynamics of a low-fitness paper strongly depends on  chance,   citation dynamics of the high-fitness paper is more deterministic.

\subsection{Divergence of citation dynamics of  the papers with the same fitness- numerical simulation}
\begin{figure}[ht]
\includegraphics*[width=0.45\textwidth]{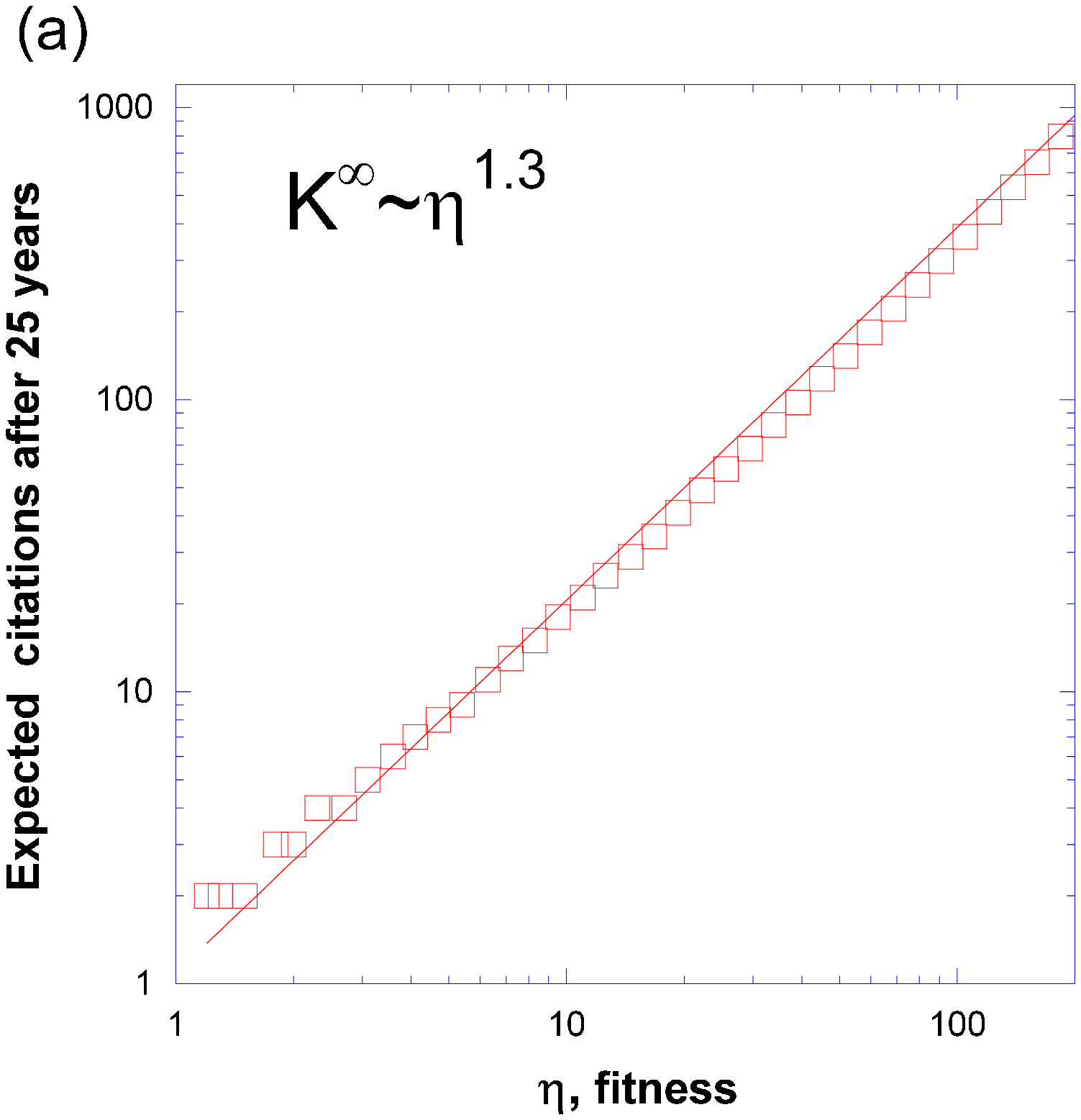}
\includegraphics*[width=0.45\textwidth]{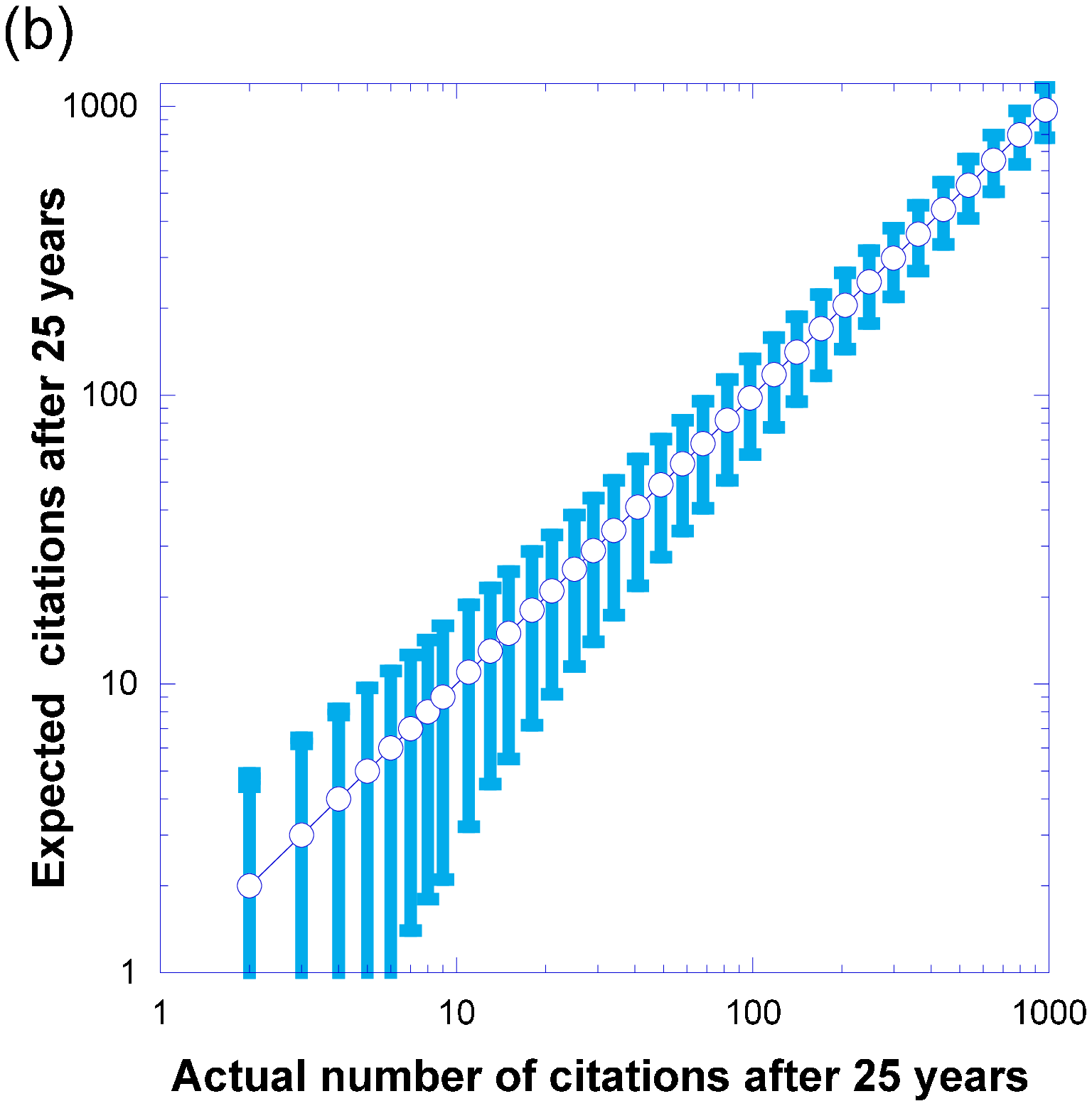}
\caption{(a) Expected number  of citations after 25 years, $K^{\infty}(\eta)$, in dependence of the paper's fitness $\eta$. Numerical simulation for 4000 papers with the same $\eta$. The simulation is  based on Eq. \ref{4-citation5} with the parameters  for Physics papers published in 1984. Continuous line shows an empirical  power-law dependence, $K^{\infty}\propto \eta^{1.3}$.  (b) Actual number of citations after 25 years versus expected number.  The error bars show the width of   the distribution, $K^{\infty}\pm std(K^{\infty})$. Citation distributions are broad for low $\eta$ and narrow for high $\eta$.
}
\label{fig:fitness-bound}
\end{figure}
In particular, Fig. \ref{fig:fitness-bound}b shows that if expected number of citations  in the long-time limit is 3, the actual number of citations  can be anything between 0 and 7; if expected number of citations is  10, the actual number  can be between  3 and 20, if the expected number is 100, the actual number can be between 50 and 130, if the expected number is 1000, the actual number  can be between 700 and 1200. If we compare two papers that garnered 3 and 20 citations in the long-time limit, they can have the same fitness $\eta$, namely, they are most probably in the same "quality" league. Two papers that garnered 700 and 1200 citations are probably in the same "quality" league, namely they can have the same fitness. But the papers that garnered 100 and 1000 citations  should have different fitness and belong  to different "quality" leagues.

\begin{figure}[ht]
\includegraphics*[width=0.45\textwidth]{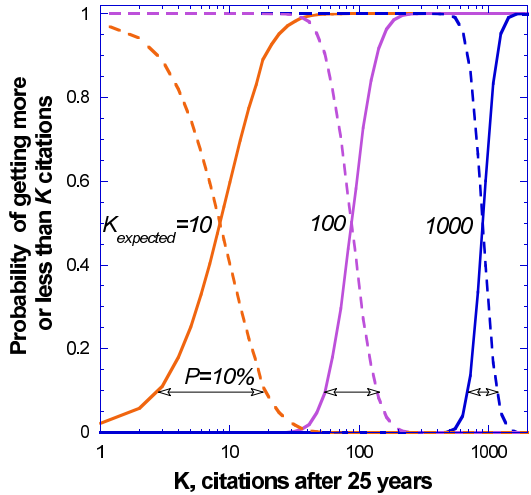}
\caption{Statistical distribution of the  number of  citations  after 25 years for the papers with the same fitness  $\eta$. Numerical simulation based on Eq. \ref{4-citation5} for 4000 papers.  The mean of the distribution, $K^{\infty}$, is indicated at each curve. Continuous lines show  cumulative probability of getting more than $K$ citations, $\int_{K}^{\infty}p(K)dK|_{\eta=const}$.  Dashed lines show complementary probability of getting less than $K$ citations, $\int_{0}^{K}p(K)dK|_{\eta=const}$. The intervals of the $10\%$ probabilities of having $K_{expected}$ citations are shown by arrows.
}
\label{fig:fitness-bound1}
\end{figure}

Figure \ref{fig:fitness-bound1} shows  $K^{\infty}$- distributions  from a slightly different perspective.  We observe that a paper which is \emph{worth} 10 citations, with 10$\%$ probability can garner less than 3 or more than 18 citations; a paper which is \emph{worth} 100 citations, with 10$\%$ probability can garner less than 54 or more than 135 citations; a paper which is \emph{worth} 1000 citations, with 10$\%$ probability can garner less than 680 or more than 1150 citations.

It should be noted that our model assumes constant fitness through the whole citation career of the paper. Similar assumption was adopted by Ref. \cite{Kong2008} in their description of Web-pages popularity and it was  justified by measurements. This assumption is reasonable for ordinary papers but not for sleeping beauties, that can be dormant for a long time and then become popular.

\subsection{Fitness estimation}
Refs. \cite{Wang2013,Stringer2008} associate fitness with the ultimate impact of the paper, namely the number of total citations in the long-time limit; Ref. \cite{Fortunato2006} determines paper fitness by ranking; Ref. \cite{Higham2019} estimate patent fitness as  a combination of attributes found through factor analysis, Ref. \cite{Simkin2007} associates fitness with the number of citations during a couple of years after publication. We define fitness slightly differently, namely, $\eta$ is  the number of direct citations in the long-time limit. Obviously, this definition cannot be a basis for prediction of future citations since it can be used only when citation career of a paper is close to completion.
\begin{figure}[ht]
\includegraphics*[width=0.45\textwidth]{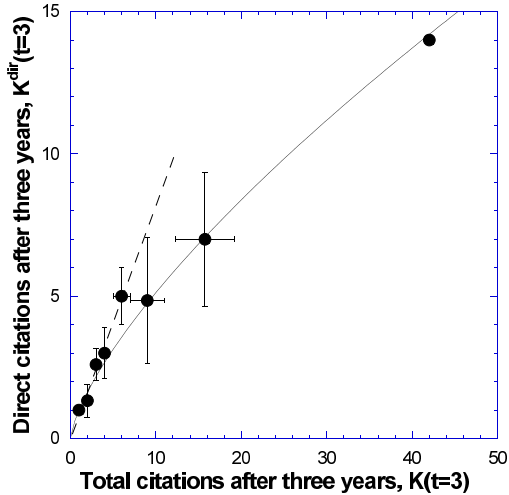}
\caption{ The relation between direct citations and total citations during first three years after publication for Physics papers published in 1984. Publication year corresponds to $t=1$. The dashed line shows linear approximation, $K^{dir}(3)=K(3)$. This approximation is good only for low-cited papers.  Red continuous line shows the empirical approximation $K^{dir}(3)=[K(3)]^{0.7}$ which shall be used for highly-cited papers. The fitness is estimated  from Eq. \ref{4-citation5} as $\eta_{j}=\frac{K_{j}^{dir}(3)}{R_{0}\sum_{1}^{3}\tilde{A}(t)}$.
}
\label{fig:fitness-bound2}
\end{figure}

To work out operational definition of fitness that can be used for prediction, we note that the fitness in our sense  is  related to initial citation rate, since at the beginning of the citation career of a paper   citations are predominantly  direct. We base our operational definition of  fitness  on the "magic of three years", well-recognized in bibliometrics.  Namely, the number of citations garnered by a paper during first 2-3 years after publication (for computer science papers this initial period is 0.5-1 year) is a basis for fitness estimation. Figure \ref{fig:fitness-bound2} shows that the relation is nonlinear (see also Fig. \ref{fig:fitness-bound}a) and from this calibration plot we estimate fitness. 

\section{Fitness estimation basing on paper's content}
We believe that $\eta_{i}$  is determined by the journal (venue),  the number of  researchers in the area, reputation of the research group, and last but not least -by the paper's  novelty, timeliness, and quality although the latter can be subjective notion.  It should be noted, however, that the paper's fitness and the number of citations  gauge not the quality of a paper but  its impact. Note, that even erroneous paper can have a great impact. On the other hand, the impact can depend on the factors unrelated to paper's content - institution, reputation of the research group, catchy title, etc.

For example, H. Brot et Y. Louzoun showed \cite{Brot1a} that   the name of the first author matters for citation count, in particular, the Physics papers whose first author's name starts from the letters A, B, C, in the long time limit have $\sim 10\%$  more citations than the papers whose list of authors starts from X, Y, Z. (May be,  success of the famous Alpher-Bethe-Gamow paper partially derives from the lucky combination of their last names?)

To further demonstrate the importance of the author's name for success of the paper, we consider  anectodal evidence based on a couple of papers. Indeed, Richard Lewontin and Jack Hubby made a landmark study in molecular evolution while collaborating in the University of Chicago. To get equal credit for their contribution, their scientific report was published  as two companion papers  with very similar titles and subjects:
 \begin{enumerate}
  \item J.L. Hubby and R.C. Lewontin, "A molecular approach to study of genic heterozygosity in natural populations. \textbf{1}. Number of alleles at different loci in drosophila pseudoobscura", Genetics 54(2), 577-594  (1966).

  \item R.C. Lewontin and J.L. Hubby,  "A molecular approach to study of genic heterozygosity in natural populations. \textbf{2.} Amount of variation and degree of heterozygocity in natural populations of  drosophila pseudoobscura", Genetics 54(2), 595-609  (1966).
\end{enumerate}

The main difference  between these two papers  is the order of authors.  By 2018, the second paper got  around 900 citations while the first paper got only around 500 citations!  This difference is explained by the fact that, when the papers were first published in 1966,  Lewontin, who was three years older than Hubby,  was  better known in the scientific community. Thus, researchers  preferred to cite the paper in which Lewontin was the first author. Eventually, Hubby became also well-known,   the paper in which he was the first author got fair credit and a large number of citations. However,  citation count of the Lewontin paper remained bigger due to impressive head start. On another hand, do  citation counts  of these two papers reflect difference in their "quality"? Our model and  Fig. \ref{fig:fitness-bound} show,  that the probability of two papers, which garnered  500 and 900 citations in the long time limit, to have the same fitness,  is $\sim10\%$. This probability is not small, hence  it is quite probable that the papers of Lewontin and Hubby are  in the same "quality" league.


\section{Timeliness of results}
One of the important criteria,  which the editor and reviewers use in their eavluation of submitted  papers, is the timeliness of results. This criterion singles out the papers that deal with a hot topic. In our parlance, the paper that focuses on hot topic has enhanced fitness as compared to the paper belonging to the mature research direction.  How one can quantify the corresponding contribution to fitness?

Suppose that at year $t_{0}$  there appeared one or several breakthrough papers which were followed by a flurry of subsequent developments. This means that a new field  (hot topic) has been born.  The number of publications in this new field starts to grow explosively and then saturates. As we have shown before \cite{Golosovsky2017}, the authors are conservative in their citing habits, and the length and the age composition of their reference lists remains more or less the same. In particular, the papers that were published in the same year constitute
$\sim 2 -3\%$ of all references, the papers published an year before constitute $\sim 8-10\%$ of all references, the papers that were published two years before also constitute $\sim 8-10\%$ of all references, etc. Thus, the papers that were published long after the onset of a new topic have big choice in choosing their references, while the papers published soon after the onset of a new topic have a very limited choice for filling their reference list and all choose the papers that were published close to the onset.  Thus, the papers that were published soon after the birth of a new field, namely, timely papers, shall have enhanced number of citations (enhanced fitness).

To put these considerations into quantitative terms, we consider a new field that appeared at time $t_{0}$. We denote the annual number of publications in this field by $N(t_{0}+t)$.  Equation \ref{4-citation5} yields the average number of direct citations that the paper in this field,  which was published in year $t_{0}+t$,  garners during  three subsequent  years,
\begin{equation}
K^{dir}(t_{0}+t,t_{0}+t+3)=\overline{\eta}(t_{0}+t)R_{0}(t_{0}+t)\sum_{1}^{3}\tilde{A}(\tau)N(t_{0}+t+\tau).
\label{fitness-model8}
\end{equation}
where $\overline{\eta}$ is the average fitness of the papers in the new  field which were published in year $t_{0}+t$, $R_{0}$ is the average reference list length of the papers published in year $t_{0}$, $\tilde{A}(\tau)$ is the aging function for citations.  Note also,  $\tilde{A}(\tau)=A(\tau)e^{(\alpha+\beta)\tau}$ where $A(\tau)$ is the aging function for references. While the aging function for citations is specific for each discipline and publication year, the aging function for references turns out surprisingly universal and almost independent of the publication year \cite{Golosovsky2017}. The reference-citation duality  \cite{Golosovsky2017} yields average fitness for the papers published in year $t_{0}+t$,
\begin{equation}
\overline{\eta}(t_{0}+t)=\frac{\sum_{1}^{3}A(\tau)\frac{N(t_{0}+t+\tau)}{N(t_{0}+t)}e^{\beta\tau}} {\sum_{1}^{3}A(\tau)e^{(\alpha+\beta)\tau}}.
\label{fitness-model9}
\end{equation}
If the new field grows with the same rate as the whole discipline, namely, $\frac{N(t_{0}+t+\tau)}{N(t_{0}+t)}=e^{\alpha\tau}$, then $\overline{\eta}$ does not depend on $t$. However, if this new field grows faster than the whole discipline, then $\overline{\eta}$ is enhanced.

\begin{figure}[ht]
\includegraphics*[width=0.37\textwidth]{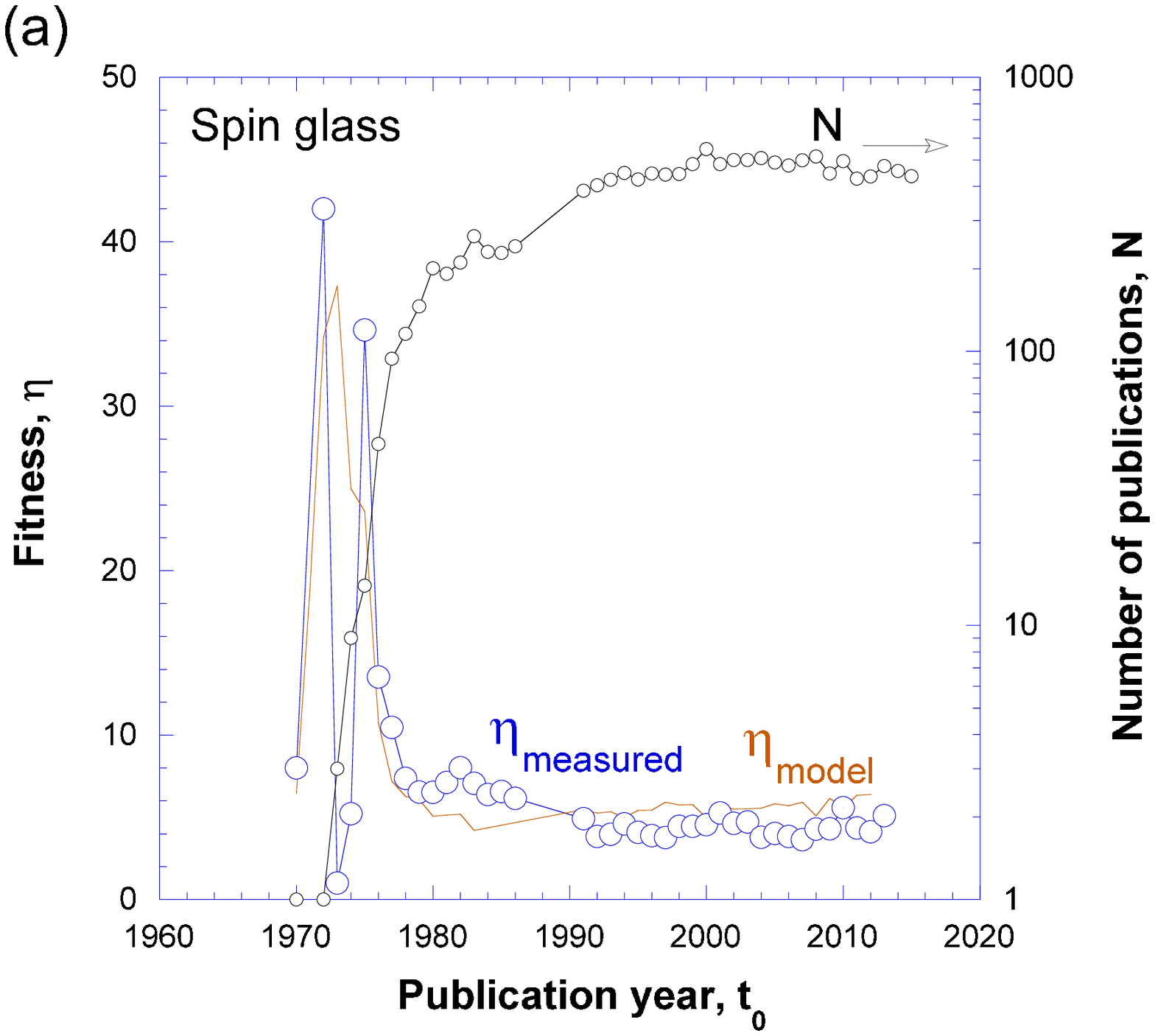}
\includegraphics*[width=0.37\textwidth]{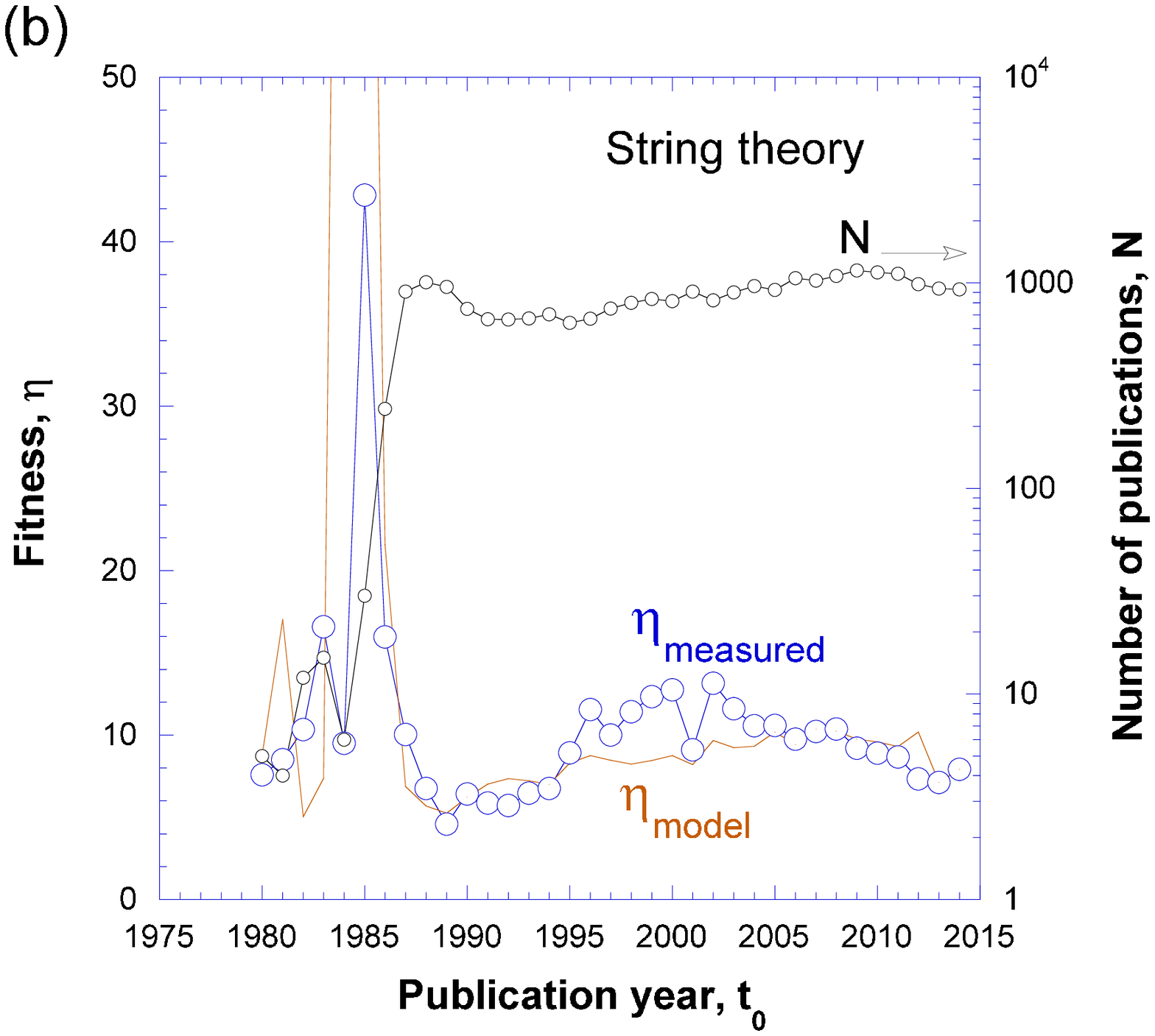}
\includegraphics*[width=0.37\textwidth]{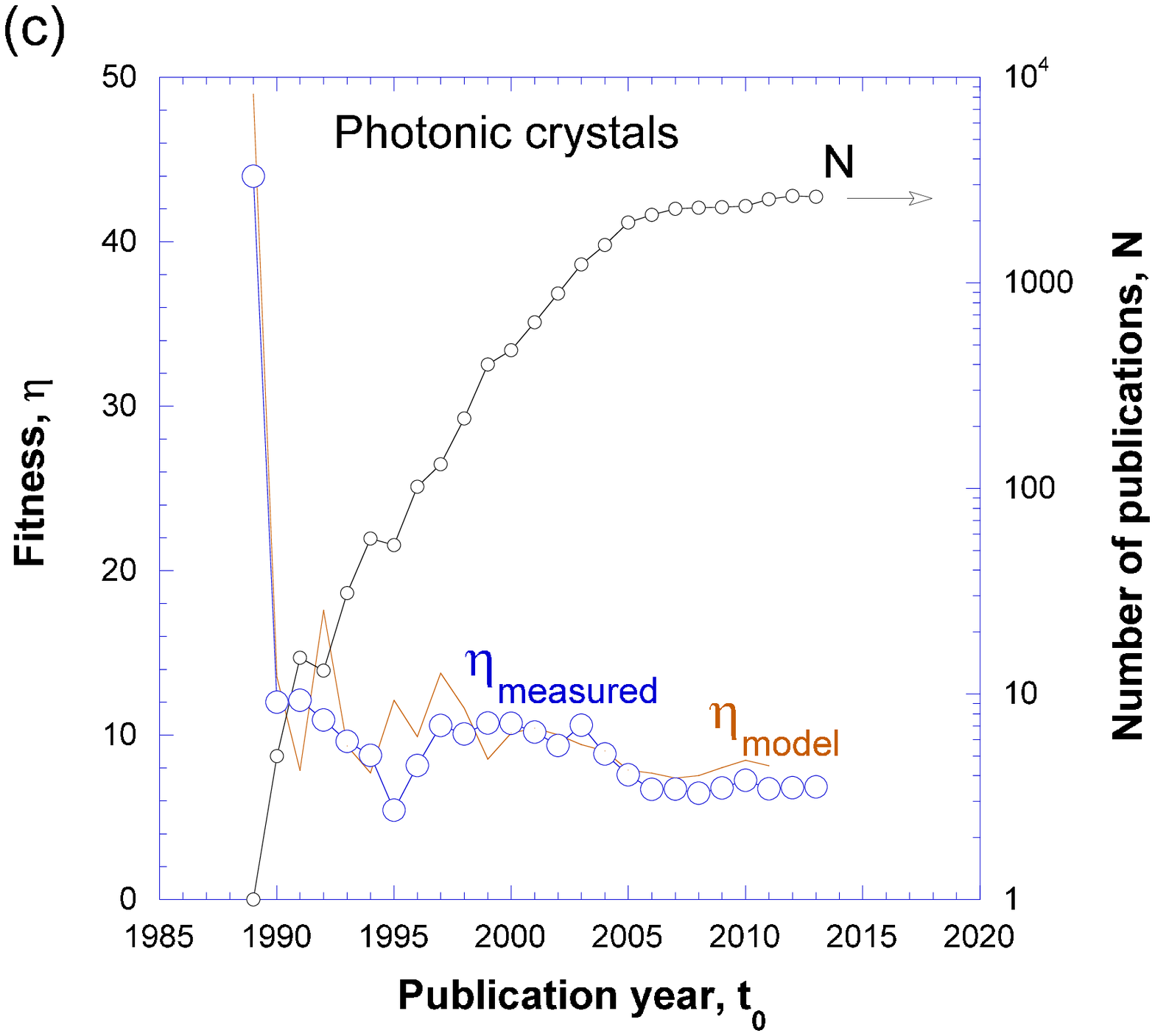}
\includegraphics*[width=0.37\textwidth]{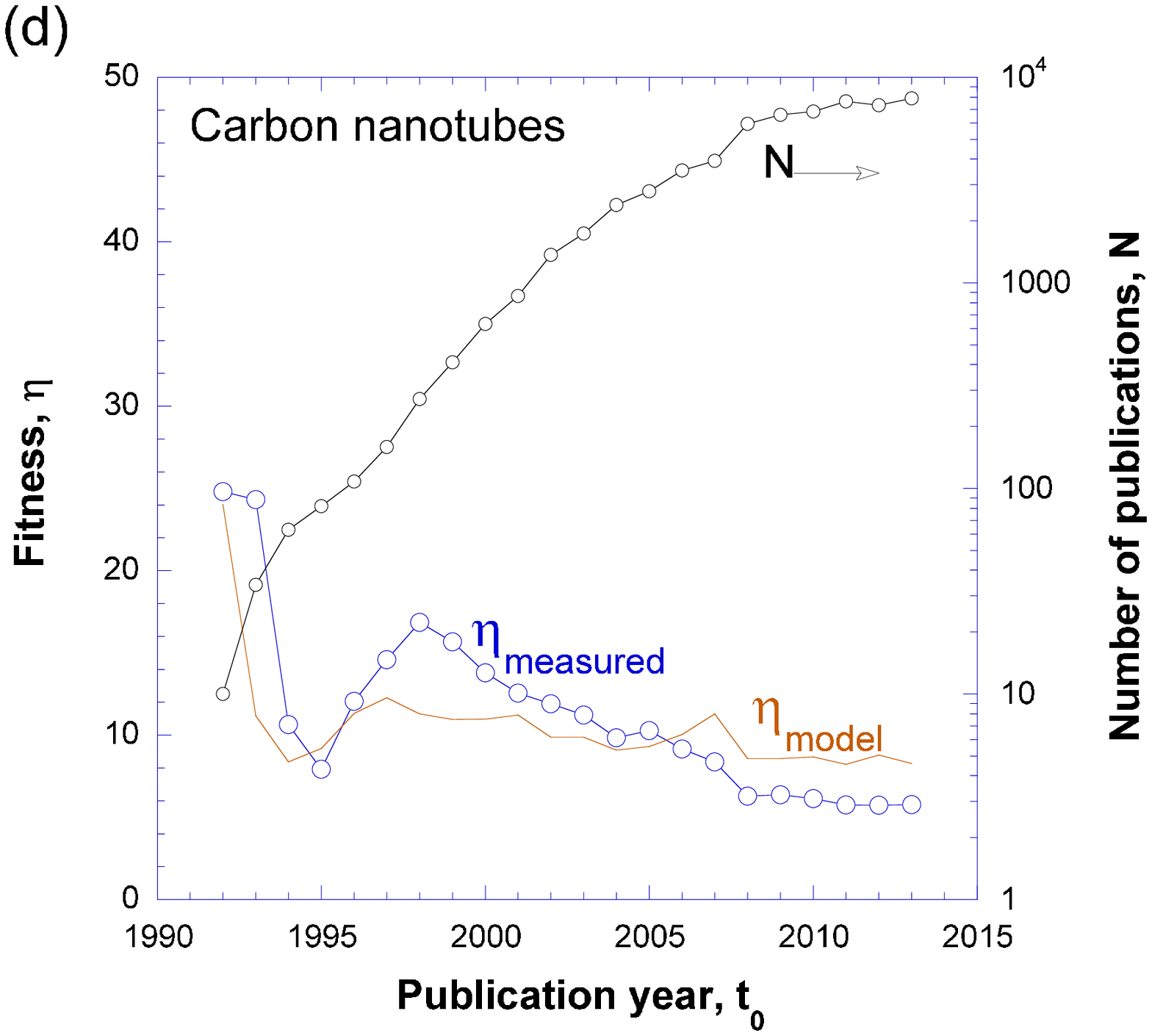}
\includegraphics*[width=0.37\textwidth]{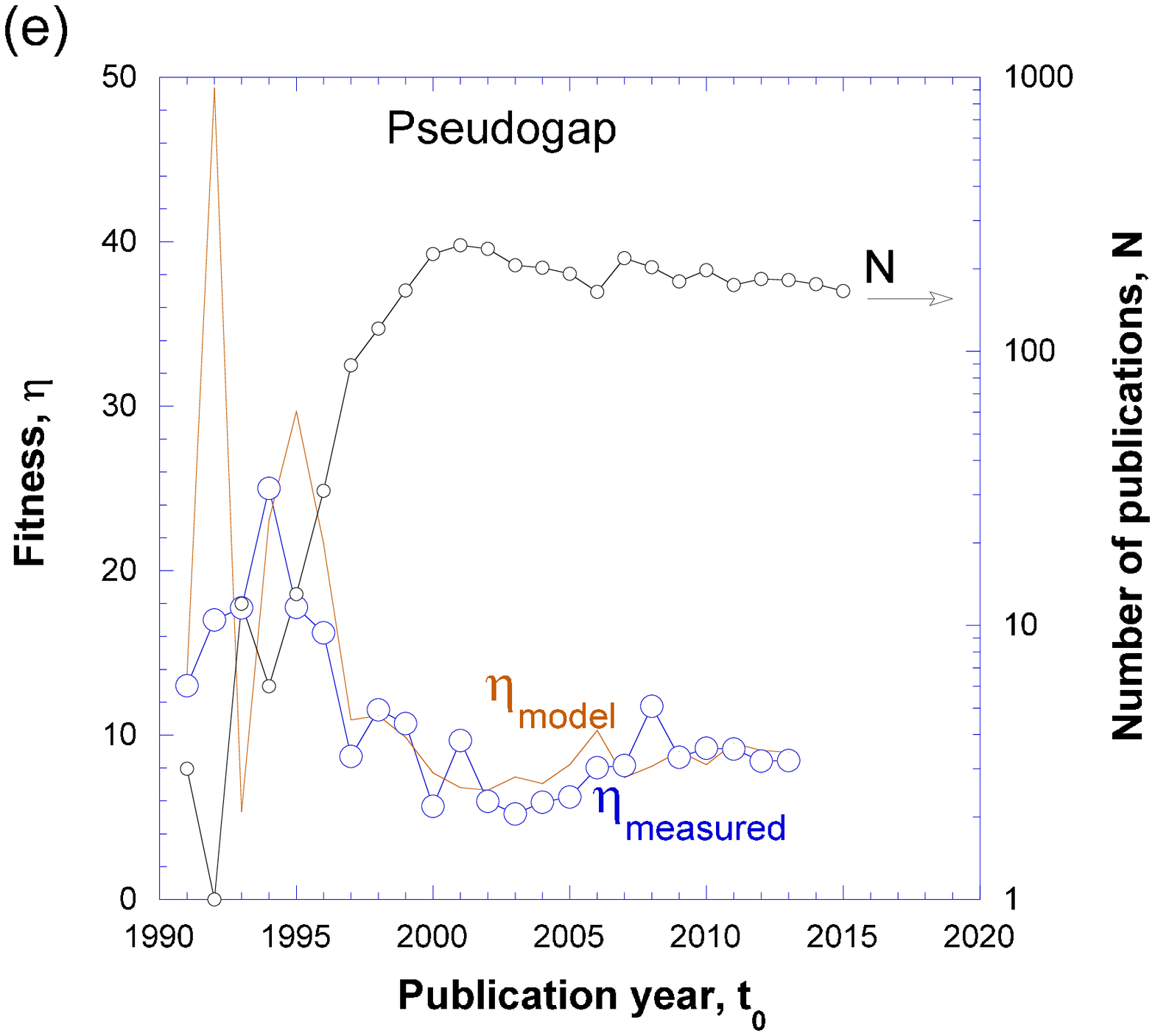}
\includegraphics*[width=0.37\textwidth]{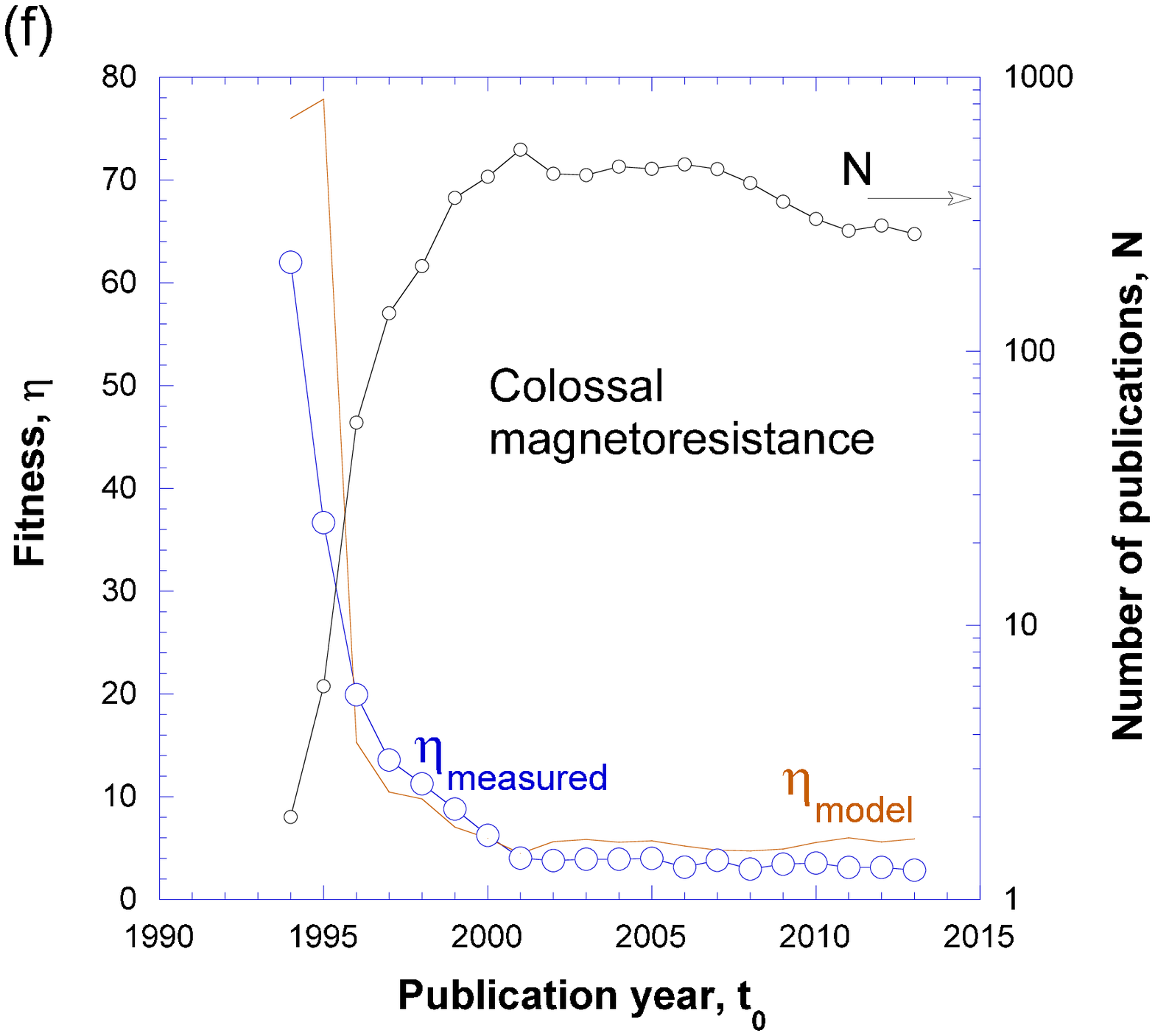}
\caption{Paper's fitness for several Physics research topics. Open black circles show number of research articles for each topic. Filled circles show our measurements of fitness  based on the number of citation garnered during first three years after publication. Blue continuous lines show model prediction based on Eq. \ref{fitness-model9}. (a) Spin glass. (b) String theory. (c) Photonic crystals \cite{photonic}.  (d) Carbon nanotubes. (e) Pseudogap. (f) Colossal magnetoresistance. 
}
\label{fig:CMR}
\end{figure}

Figure \ref{fig:CMR} illustrates these considerations. We know that a hot topic usually appears abruptly and can be identified through a burst of citations and publications \cite{Leydesdorff2018}. We choose several such research areas in Physics with well-defined onset $t_{0}$, with some of these areas the author of this book has had personal experience. Using Web of Science,  we found all papers belonging to each of these topics, that were published in year $t_{0}+t$.  For each $t$, we measured  annual number of papers and statistical distribution of the  number of citations garnered  by them during first three years after publication. Then we determined the mean and the width of these distributions.    Using Eq. \ref{fitness-model9} and Fig. \ref{fig:fitness-bound2}, we found the  average fitness of the papers  in each topic published in year $t_{0}+t$, basing on the mean of the distribution.   On another hand, we estimated this fitness using Eq. \ref{fitness-model9}. Figure \ref{fig:CMR} shows that the model prediction based on Eq. \ref{fitness-model9} captures our measurements perfectly well.

Figure \ref{fig:CMR}  implies that any paper published soon after the new topic appeared, has a good head start and this quantifies the "first mover advantage" introduced by Newman \cite{Newman2009}. However, this does not mean that the papers published long after the onset of a hot topic doomed to be undercited. In fact, Fig. \ref{fig:CMR}  shows only the mean of the fitness distribution for each year. The actual fitness distribution is very wide and its width is comparable to the mean. Hence, at each moment after the onset of a hot topic  there are many papers whose fitness considerably exceeds the average one.
\section{Discussion}
We showed here that our stochastic model of citation dynamics can be a basis for predicting citation trajectory of papers. This model shall be compared to the physics-inspired predictive model developed by Wang, Song, and Barabasi  \cite{Wang2013}.  Pham, Sheridan, and Shimodaira \cite{Pham2015,Pham2016} developed a software package  based on this model and demonstrated that it is a valid predictive tool. This model includes three paper-specific parameters: fitness $\eta$, immediacy $\mu$, and $\sigma$. To determine these parameters, one needs to measure initial citation trajectory of a paper, 2-3 years are not enough. As a predictive tool, this model works best for the highly-cited papers. Although this deterministic model  predicts citation trajectory of a paper, it  cannot specify probabilistic margins of the prediction. On the contrary, our probabilistic model includes only one paper-specific parameter- fitness, it does provide probabilistic margins of the future citation count. However, our model works better with ordinary papers and does not predict well citation trajectories of the highly-cited papers. Thus, our model is complementary to that of Ref. \cite{Wang2013}.

What are its possible applications? We believe that our model can be used for forecasting the five-year  journal impact factor. The papers published in one year in one journal represent more or less homogeneous set of papers, hence predicting the mean number of citations for this set is more reliable than predicting citation trajectory of a single paper. On another hand, our model can give probabilistic margins of such prediction.

Another application can be the early identification of the breakthrough papers. So far, this was done by analyzing diversity and age structure of the reference list of papers \cite{Uzzi2013,Mukherjee2017}, diversity and interdisciplinarity of paper's content \cite{Ponomarev2014a}, or through identification of the atypical citation trajectory, corresponding to sleeping beauties \cite{Ke2015}. An important question is how soon can we identify such rising star? Obviously, if the paper (or patent) gets more citations than what is expected from the ordinary paper published in the same year and in the same journal, then this is a candidate to be a breakthrough paper \cite{Mariani2018}. On another hand, the deviation from the ordinary citation trajectory may be accidental. Our model can make an estimate of the probability of the  enhanced citation count in order to judge whether it occurred by chance or not.
\pagebreak
\bibliography{reference_master_2019_new}
\end{document}